\documentclass{emulateapj}  

\usepackage{graphicx} 
\newcommand{\beq}{\begin{equation}} 
\newcommand{\eeq}{\end{equation}} 
\newcommand{\beqa}{\begin{eqnarray}} 
\newcommand{\eeqa}{\end{eqnarray}} 
\def\nn{\nonumber} 
\def\l({\left[} 
\def\r){\right]} 
\def\ltap{\raisebox{-.55ex}{\rlap{$\sim$}} \raisebox{.4ex}{$<$}} 
\def\gtap{\raisebox{-.55ex}{\rlap{$\sim$}} \raisebox{.4ex}{$>$}} 
\def\gsim{\mathrel{\gtap}} 
\def\lsim{\mathrel{\ltap}} 
\def\gr{$\gamma$-ray} 
\def\m87{M~87}

\usepackage{color} 
\definecolor{red}{rgb}{0.7,0,0} 
\definecolor{blue}{rgb}{0,0,0.7}

\shorttitle{TeV emission from \m87} 
\shortauthors{Neronov \& Aharonian} 
 
\begin{document} 
\title{Production of TeV gamma-radiation   
in the vicinity  of the \\  
supermassive black hole in the giant radiogalaxy M87} 
\author{A.Neronov} 
\affil{{\it INTEGRAL} Science Data Center,\\
16 ch. d'Ecogia, CH-1290, Versoix, Switzerland\\
and\\
Geneva Observatory, University of Geneva\\
51 ch. des Maillettes, CH-1290 Sauverny, Switzerland}
\email{andrii.neronov@obs.unige.ch}
\and
\author{Felix A. Aharonian}
\affil{Dublin Institute for Advanced Studies, 5 Merrion Square,
Dublin 2, Ireland \\ 
and \\
Max Planck Institut f\"ur Kernphysik, Saupfercheckweg 1, 69117 Heidelberg, Germany} 
\begin{abstract}
Although the giant radiogalaxy \m87\ harbors many distinct  regions of 
broad-band nonthermal emission, the recently  reported fast variability of TeV 
\gr s from \m87\  on a timescale of days strongly constrains the range of  
speculations concerning the possible sites and scenarios of particle  
acceleration responsible for the  observed TeV emission. A   natural production
site  of this radiation  is the immediate vicinity of the   central
supermassive  mass black hole (BH).   Because of the low bolometric 
luminosity,  the nucleus of \m87\ can be effectively transparent for \gr s  up to 
energy of 10~TeV, which makes this source an ideal laboratory for study of  
particle acceleration processes close to the  BH event horizon. We critically 
analyse different possible radiation mechanisms in this region, and  argue that
the observed  very high-energy \gr\ emission  can be explained by the inverse
Compton  emission of  ultrarelativistic electron-positron pairs  produced 
through the  development of an electromagnetic cascade  in the BH
magnetosphere. We  demonstrate, through detailed numerical calculations    of
acceleration  and  radiation of electrons  in the magnetospheric vacuum gap,
that this   ``pulsar magnetosphere like'' scenario can satisfactorily explain
the main  properties of TeV gamma-ray emission of \m87.  
\end{abstract} 
 
                   
\section{Introduction} 
\label{intrd} 

\m87, a nearby giant radio galaxy, located at a distance of $d\simeq 16$~Mpc 
\citep{tonry91},  hosts one of the most massive ($M\simeq 3\times 
10^9M_\odot$)   black holes (BH) in the nearby Universe \citep{marconi}. \m87 
contains  a famous kpc-scale jet the high-resolution images of which detected 
at radio, optical and X-ray wavelengths show several prominent structures. 
Nonthermal processes play important, if not the dominant, role across the 
entire jet. The apparent  synchrotron origin of the detected nonthermal 
emission,  which extends from radio to X-ray bands,  implies effective 
acceleration of  electrons to multi-TeV energies. One may expect that protons 
and nuclei, which do not suffer radiative losses, are accelerated to much 
higher energies. Both the inner (sub-parsec) and large (kpc) scale parts of 
the jet of \m87\ are possible sites of  acceleration of protons to extremely 
high energies  ($E \sim 10^{20}$~eV). Therefore   production of  gamma-rays  in
different segments of the jet due to  electromagnetic or  hadronic processes is
not only possible, but, in fact,  unavoidable. The jet  of \m87\ is observed at
large angle, $\sim 20^\circ$ \citep{biretta99}. Therefore, unlike  blazars, we
do not expect a  strong Doppler boosting of the \gr\ flux.  On the  other hand,
the nearby location of \m87\ compensates this disadvantage (compared to 
blazars)  and makes several prominent  knots and hot spots of the jet 
as potentially detectable  TeV \gr\ emitters. 

In this regard, the  discovery of a TeV \gr\ signal from \m87\ by the HEGRA 
array of Cherenkov telescopes \citep{HEGRA} and its confirmation by the HESS 
array of telescopes \citep{HESS},  was not a big surprise, especially given the 
rather modest apparent TeV \gr\  luminosity  (few times $10^{40} \ \rm 
erg/cm^2 s$).  Several electronic and hadronic models have been  suggested 
for explanation of  TeV \gr\ emission of  M87. The suggested sites of TeV \gr\ 
production range from large scale  structures of the kpc jet \citep{Lukasz} to  
a compact peculiar hot spot (the so-called  \textit{HST}-1 knot) at a distance
100 pc    along the jet \citep{HST-1} and  inner (sub-parsec) parts of the jet 
\citep{Markos,Anita}. 

While gamma-ray observations cannot provide  images with an adequate
resolution  which would  allow localisation of sites of \gr\ production,  the
variability  studies can discard or effectively  constrain the suggested
models.  

The continuous monitoring of \m87\ with the HESS telescope array during the 
period 2003-2006 not only revealed statistically significant fluctuations of 
the TeV flux on a yearly basis, but, more excitingly, an evidence of fast 
variability on timescales of $\Delta t \sim 2$ days was found in the 2005
dataset  \citep{HESS}. This requires a very compact region with a characteristic
linear  size $R \leq \Delta t \delta_{\rm j} \simeq 5 \times 10^{15}
\delta_{\rm j} \  \rm cm$, where $\delta_{\rm j}$ is the Doppler factor of the
relativistically  moving source (throughout the paper we will use the system of
units in which  the speed of light $c=1$).  Note that since the mass of the BH
in \m87\ is well  established \citep{marconi}, the Schwarzschild radius is
estimated quite  accurately $R_{\rm  Schw}=2GM\simeq 10^{15}\left[M/3\times
10^9M_\odot\right]$~cm ($G$ is the  gravitational constant). The expected
minimal variability time scale (light  crossing time of the black hole) for a
non-rotating BH is   $T_{\rm ls}=2R_{\rm Schw}/\simeq 6\times 10^4$~s$\simeq
1$~day, while  it is two times less for the maximally rotating Kerr BH.  
The observed  variability time scale of TeV emission of $\sim
2$~days indicates that the  emission originates within the last stable orbit of rotation around the black hole,
unless the radiation is  produced in relativistically moving outflow with a
Doppler factor $\geq 10$.  Although there are sound arguments against \m87\
being a blazar \citep{Fabian},  one cannot in principle exclude that at the base
of its formation (close to the  BH), where \gr s are produced, the jet is
pointed to us, and only later, it  deviates from our line of sight. In this
regard, both leptonic \citep{Markos} and hadronic \citep{Anita} models suggested
for TeV radiation of \m87\ cannot be  discarded.  Note however, that both
models predict rather steep energy spectra  contrary to the observed hard \gr\
spectrum extending to $E \ge 10 \ \rm TeV$ \citep{HESS}. Whether these models
are flexible enough to reproduce the observed  TeV \gr\ spectrum by tuning the
relevant model parameters and introducing  additional assumptions, should show
further detailed theoretical studies. 

Finally, one should mention that formally one may assume that  \gr s are 
produced in a compact region  far from the central engine. In this regard, the 
famous \textit{HST}-1 knot has certain attractive features which make this 
compact structure  a potential site of particle acceleration and \gr\ 
production \citep{HST-1}.  Although the favored size of this structure  is in 
the range between 0.1 and 1 pc, and thus contradicts the  observed TeV \gr\ 
variability, the lack of robust lower  limits on the size of \textit{HST}-1 
leaves the \gr\ production in   this peculiar knot as a possible option. 
Nevertheless one should note that the location of \textit{HST}-1 at a distance 
of  100 pc from the central BH requires (almost) unrealistically tight 
collimation of the jet. 

In this paper we assume that the TeV \gr\  production takes place close to the 
event horizon of the central supermassive BH, and show that the acceleration 
of  electrons in a vacuum  gap in BH magnetosphere can explain the  general 
characteristics of the   TeV $\gamma$-ray emission observed from M87.   

Such mechanism of \gr\ emission from the vicinity of a black hole  is a close
analog of the mechanism of pulsed  \gr\ emission from the vicinity  of neutron
stars in pulsars. The similarity between the electrodynamics of the pulsar and
black hole magnetospheres was  discussed in the seminal paper of
\citet{blandford77} in the context of pair production and energy transfer of
rotational energy of a  black hole through the Poynting flux. The mechanisms
of  emission of high energy \gr s from the direct vicinity of black holes  by
electrons and protons accelerated in the electric field of vacuum gaps 
were discussed by \citet{blandford77} (electron curvature radiation),
\citet{beskin92} (inverse Compton scattering by electrons) and
\citet{levinson} (proton curvature radiation).

\section{Internal absorption of $\gamma$-rays} 
\label{sec:data} 

\begin{figure} 
\includegraphics[width=\linewidth]{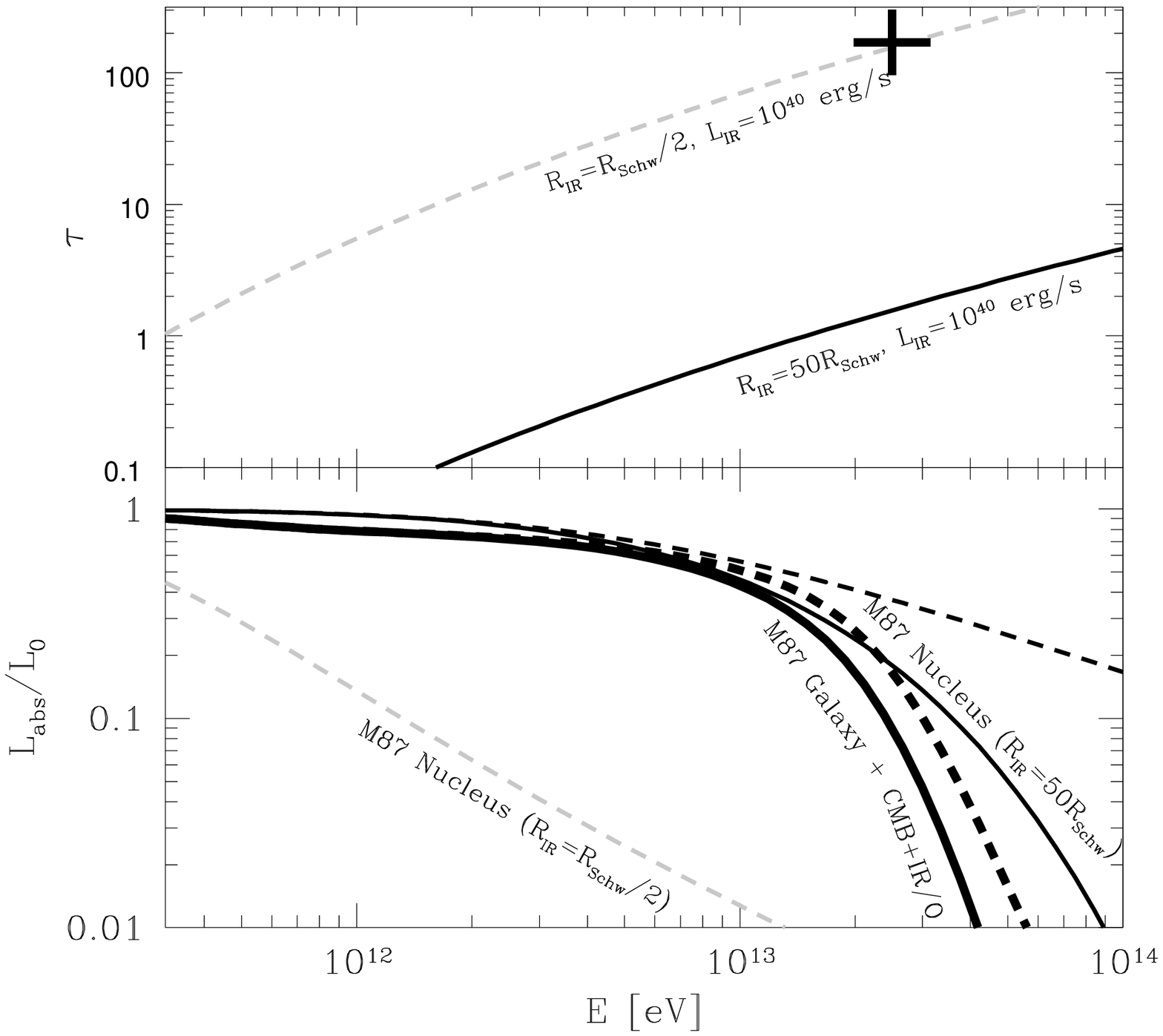} 
\caption{  Top panel: optical depth for \gr s produced in the vicinity  of the
black hole in the cases of the infrared source of size $50 R_{\rm Schw}$ (black
solid curve) and $R_{\rm Schw}/2$ (grey dashed curve).  The estimate of
\citep{cheung07} is shown by a cross. The spectrum of the infrared background
in the source is the one shown by red dashed curve in Fig. \ref{fig:SED}).
Bottom panel: attenuation of $\gamma$-rays in \m87\ due to  photon-photon pair 
production.  The internal absorption (thin solid curve) is   dominated by
interactions with the infrared radiation of the   compact source in the core of
\m87.  The external absorption due to the   interaction with the diffuse
radiation fields within the   elliptical galaxy \m87, the 2.7 K CMBR and
the diffuse   extragalactic infrared  background  photons leads to a further
suppression of the \gr\ flux, shown by thick solid curve. Attenuation 
in the case when the \gr\ emission is distributed
throughout the infrared source is shown by thin (intrinsic absorption in the
source) and thick (modification during the propagation through the galactic and
extragalactic background light) dashed curves. Dashed grey curve shows the
absorption of the \gr\ flux in the case of a "maximally compact" infrared
source of the size $R_{\rm IR}=R_{\rm Schw}/2$ (tee text). } 
\label{fig:tau} 
\end{figure} 
 
The observed infrared luminosity of the nucleus of \m87,  $\nu L_\nu\sim   
10^{40\div 41}$~erg/s \citep{perlman,whysong04} is  6.5 orders of magnitude
lower than the  Eddington luminosity of a $3\times 10^{9}M_\odot$ BH.  In this
regard the BH of  in \m87\ is similar to the supermassive BH in the center of
our   galaxy.  In both cases  the low bolometric luminosity of the nucleus
makes the  "central engine" of activity, i.e. the vicinity of the event horizon
of the  supermassive BH, transparent to the very high energy (VHE) \gr s
\citep{AhNer2005}. 

In the isotropic field of background photons, the cross-section of
photon-photon  pair  production depends on the product of energies of
colliding photons, $s=E \epsilon  /m_{\rm e}^2$. Starting from the threshold at
$s=1$, the cross-section   $\sigma_{\rm  \gamma \gamma}$ rapidly increases
achieving the  maximum  $\sigma_0  \approx \sigma_{\rm T}/5 \simeq 1.3 \times
10^{-25} \ \rm cm^2$ at  $s \approx 4$,  and then decreases as $s^{-1} {\rm ln}
s$. Because of relatively narrow  distribution of $\sigma_{\rm \gamma
\gamma}(s)$, gamma-rays interact  most  effectively with the infrared
background photons of  energy  
\beq 
\epsilon \approx 1 (E /1 \ \rm TeV)^{-1} \  eV \ . 
\label{epsilon} 
\eeq 
Thus the optical depth for a gamma-ray of energy $E$ produced in the center 
of the infrared source of the size $R_{\rm IR}$ and 
the luminosity $L_{\rm IR}$ at energy given by Eq.(\ref{epsilon}) 
can be written in the form
\beqa 
\label{opt} 
&&\tau (E,R_{\rm IR}) =\frac{L_{\rm IR} \sigma_{\gamma\gamma}}{4 \pi R_{IR}\epsilon}
\simeq \\ &&
0.1 \left[\frac{L_{\rm IR}(1 [E/1{\rm TeV}]^{-1}{\rm eV})}{3\times 10^{40}\mbox{ erg/s}}
\right] 
\left[\frac{R_{\rm IR}}{50R_{\rm Schw}}\right]^{-1} \left[\frac{E}{1\mbox{ 
TeV}}\right] \ . \nonumber
\eeqa
The dependence of the optical depth on gamma-ray energy is determined by the
spectral form of background radiation $n(\epsilon)=L_{\rm IR}(\epsilon)/(4\pi
R\epsilon)$. In particular, in the case of power-law spectrum with photon index
$\Gamma$ ($n(\epsilon) \propto \epsilon^{-\Gamma}$, or $L_{\rm IR} (\epsilon)
\propto \epsilon^{-\Gamma+1}$,  one has $\tau(E,R_{\rm IR}) \propto
E^{\Gamma-1}$. Accurate numerical calculations of the optical depth for the
spectral energy distribution of the compact infrared source in the nucleus of
M87 (see Fig.\ref{fig:SED}, Section 5) and normalized to the source size
$R_{\rm IR}=50 R_{\rm Schw}$ is shown by black
solid curve in the upper panel of Fig.\ref{fig:tau}. Since  $\tau(E, R_{\rm
IR}) \propto R_{\rm IR}^{-1}$,  the optical depth does not exceed 1 even at the
highest detected energies of gamma-rays of about 10 TeV, provided that infrared
source is larger than $50 R_{\rm Schw}$.  This is demonstrated in the lower
panel of Fig. \ref{fig:tau}. Note that the recent claim by \citet{cheung07} 
that the central region of M87 is excluded as  a site of the TeV emission
because of absorption of \gr s, is misleading. The authors obtained very large
optical depth relevant to the energy $\simeq 25 \ \rm TeV$ and assuming an
extremely compact infrared source with a linear size of $R=R_{\rm g}=R_{\rm
Schw}/2$ (the estimate of the optical depth by \citet{cheung07} is shown by a
cross in the upper panel of Fig. \ref{fig:tau}, and the dependence of the
optical depth on the \gr\ energy is shown by the grey dashed curve).  Although
formally one cannot rule out  such a compact size of the  IR source, a
significantly  larger size cannot be {\it a priory} excluded either. Moreover,
there are not special reasons to assume that the infrared source is located
very close to the event horizon.
 
For the nucleus of \m87, there are no direct measurements of the size of the 
infrared source. Observations in the microwave band at 43~GHz suggest  that the
size of the source at the mm wavelength  is limited by $5\times 10^{16}$~cm, or 
approximately $50R_{\rm Schw}$. Lower angular resolution of the infrared
telescopes does not allow us to constrain (or marginally resolve, 
see \citet{perlman}) the size of the  nuclear source  to $\le 10$~pc
\citep{whysong04}.  However, even assuming that the size of the infrared source
is comparable to the size of the microwave source, one finds from the above
estimate that the nucleus can be transparent  to \gr s with energies up to
$\sim 10$~TeV.  

Even in the case of the "maximally compact" infrared source of the size $R_{\rm
IR}\sim R_{\rm Schw}/2$, the source is partially transparent for \gr s.  
In spite of the fact that  the  optical  depth for \gr s produced in the
center of the infrared source 
\gr s becomes very large at energies $E>10$~TeV 
(see the dashed grey curve on the upper panel of Fig.
\ref{fig:tau}), there is no catastrophic absorption of  multi-TeV 
$\gamma$-rays.  The reason is that in this case the  source(s) of \gr\
photons are distributed throughout  the infrared source and  the thickness $H$
of the transparent surface layer of  the source is determined from the
condition $\tau(E,H)\simeq1$. Assuming a  homogeneous \gr\ source, one can find
that the luminosity of the last  transparent layer ($\tau\le 1$) is only
moderately, by a factor of $H/R_{\rm IR}\sim  \tau(E,R_{\rm IR})$ (rather than
by a factor of $\exp(-\tau(E,R_{\rm IR})$)  lower than the total luminosity of
the source.   The attenuation of the \gr\ flux in the case of the \gr\ source distributed throughout the infrared source is shown by dashed curves in the lower panel of Fig. \ref{fig:tau} 
for both cases of $50R_{\rm Schw}$ (black) and $R_{\rm Schw}/2$ (grey)
size of the infrared source. We assume the suppression factor $(1+\tau(E,R_{\rm IR}))^{-1}$ which is an interpolation between no supression in the $\tau=0$ limit and $1/\tau$ suppression in the $\tau\gg 1$ limit. 

The \gr s after they escape the nucleus are further attenuated due the pair 
production in the radiation fields both inside and outside  the elliptical 
galaxy  M87. The spectrum of emission from the galactic bulge of \m87\
sharply peaks  at photon energies around $\epsilon_{\rm bulge}\simeq 1$~eV.
Interactions of   nuclear \gr s with the photon background in \m87\ galaxy
should therefore lead  to maximum absorption  at \gr\ energies around
$E_\gamma\simeq 1$~TeV. The   column  density of infrared/optical photons in
the bulge of the size $R_{\rm bulge}$ and  luminosity $L_{\rm bulge}$ along the
line of  sight is estimated  as 
\begin{eqnarray} 
&&N_{ph}(1\mbox{ eV})= \int_0^{R_{\rm bulge}} n_{ph}(r)dr\simeq \\&& 5\times 10^{23} 
\left[\frac{L_{\rm bulge}}{10^{45}\mbox{ erg/s}} 
\right]\left[\frac{1\mbox{ kpc}}{R_{\rm bulge}}\right]\mbox{ cm}^{-2}\nonumber \ , 
\end{eqnarray} 
which allows to estimate the optical depth of gamma-rays at 1 TeV:   
\begin{eqnarray} 
&&\tau_{\rm\m87}(1\mbox{ TeV})=\sigma_{0}N_{ph}(1\mbox{ eV})\simeq  
\\&& 
0.08\left[\frac{L_{\rm bulge}}{10^{45}\mbox{ erg/s}} 
\right]\left[\frac{R_{\rm bulge}}{1\mbox{ kpc}}\right]^{-1}  \nonumber 
\end{eqnarray} 
Propagation of \gr s through the cosmic microwave and infrared backgrounds
over   the way from \m87\ to the observer leads to further absorption of the
highest  energy quanta. Photons with energies above $10^{15}$~eV are completely
absorbed  due to interactions with the 2.7~K microwave background (the minimal
propagation distance  is $\sim 8$~kpc). Photons with energies above
100~TeV interact most  efficiently with the far-infrared background photons
whose density is some   3 orders of magnitude lower than the density of the
microwave photons. However,  the mean free path of $E \geq 10$~TeV \gr s
interacting with far-infrared background,  is still shorter than the distance
to \m87\ (16~Mpc). 
 
\section{Physical parameters of the central engine} 
 
The high resolution observations of the nucleus of \m87\ in X-rays with  the
{\it Chandra}  observatory provide an important information about the 
accretion  onto the supermassive BH \citep{dimatteo03}. In particular, they give
an estimate of the electron density of  plasma with a
temperature $kT \sim 1$~keV,  
 \begin{equation} 
 n_e\simeq 0.1\mbox{~cm}^{-3} 
 \end{equation} 
at  the distance of the order of Bondi accretion   radius, $R_{\rm Bondi}\simeq
5\times 10^5R_{\rm Schw}$.  The corresponding accretion rate inferred from this
estimate is  
\begin{equation} 
\dot M_{\rm Bondi}\simeq 0.1 M_\odot\mbox{~yr}^{-1} \ . 
\end{equation} 
Interestingly, the observed bolometric  luminosity of the nucleus of \m87\ is
4  orders of magnitude below the expected nuclear luminosity  corresponding to
this  accretion rate 
\begin{equation} 
\label{bondi} 
L_{\rm Bondi}\simeq 10^{45}\left[\frac{\eta}{0.1}\right]\left[\frac{\dot M_{\rm 
Bondi}}{0.1M_{\odot}/\mbox{yr}}\right]\mbox{ erg/s}.  
\end{equation} 
where $\eta\sim 0.1$ is the efficiency of conversion of the rest energy of 
accreting particles into radiation. This indicates that either the
accretion  proceeds in a radiatively inefficient way, or the actual accretion
rate is  still lower than the one inferred from X-ray observations. 
 
To estimate the plasma density close to the  event horizon of the black hole, 
one has to assume a certain radial density profile, $n(r)\sim r^{-\gamma}$. 
Depending on the model of accretion flow,  the index  $\gamma$ can  vary
between $1/2$ (this value is, in fact, a lower limit  which can  be
realized  for collisionless motions of individual particles in  the central
gravitational field) and $3/2$.  The lack of information about the  accretion
regime leads to a significant  uncertainty of the plasma density near  the
event horizon,  
\begin{equation} 
10^{1.5}\mbox{ cm}^{-3}<n<n_{\rm max}\simeq 10^{6.5}\mbox{ cm}^{-3} \ . 
\end{equation}  
 
Regardless of the uncertainty of this estimate, one may  conclude that the 
strength of magnetic field in the  vicinity of the BH can not be very  high.
Indeed,  assuming that the magnetic field is generated by 
the accreting matter,  one
can find that the energy density of magnetic field can not exceed the density of the total kinetic energy stored in  the particles of the accretion flow. 
In this case even if the accreting matter moves with relativistic speed,
the estimate of maximal possible magnetic field is 
(assuming that the matter density is  $n\sim n_{\rm max}$) 
\begin{equation} 
\label{beq} 
B_{\rm eq}\simeq (8\pi n_{\rm max}  m_e)^{1/2}\sim 10\mbox{ G.}  
\end{equation} 
Thus, particle acceleration close to the BH horizon proceeds in the  relatively
low-density and low-magnetic field environment which significantly  limits the
range of possible mechanisms of  VHE
\gr\ emission.  Even for the maximally possible acceleration rate, 
$dE/dt\simeq eB_{\rm eq}$,  one can find that particles accelerated in a region
of a  linear size of about the Schwarzschild  radius can not reach energies
higher  than 
\begin{equation} 
\label{emax} 
E_{\rm max}\le eB_{\rm eq}R_{\rm Schw}\simeq 10^{18}\mbox{ eV,} 
\end{equation} 
unless the magnetic field is significantly larger than the equipartition
estimate, given by Eq. (\ref{beq}).
  
\section{Gamma-ray emission from accelerated protons.} 
 
Protons accelerated near the BH horizon can produce   \gr\ emission in the VHE
band through several radiation mechanisms.  For example, TeV emission can be
synchrotron or curvature \gr\ emission which  accompanies  proton acceleration
\citep{levinson,Aharonianetal2002,neronovtinyakov}.  The  energy loss time for
protons emitting  synchrotron radiation at the  energy $\epsilon_{{\rm
synch},p}$ is short enough to explain the observed day-scale  variability of
the signal,  
\begin{equation} 
t_{{\rm synch},p}\simeq 2.5\left[\frac{B}{10\mbox{ 
G}}\right]^{-3/2}\left[\frac{\epsilon_{{\rm synch},p}}{1\mbox{ TeV}}\right]^{-1/2}\mbox{ 
d,} 
\end{equation} 
However, the energy of synchrotron and/or curvature photons produced by  
protons accelerated to the energy $E_{\rm max}$, given by Eq. (\ref{emax})
is too low to explain the emission at 1-10~TeV,   
\begin{equation} 
\epsilon_{{\rm synch},p}\simeq 
0.1\left[\frac{B}{10\mbox{ G}}\right]\left[\frac{E_p}{10^{18}\mbox{ eV}} \right]^2\mbox{ GeV} 
\end{equation} 
and 
\begin{equation} 
\label{curvp} 
\epsilon_{{\rm curv},p}\simeq 0.01\left[\frac{E_p}{10^{18}\mbox{ eV}} 
\right]^3\left[\frac{R_{\rm Schw}}{R_{\rm curv}}\right]\mbox{~GeV} 
\end{equation} 
(assuming that typical curvature radius of proton trajectories is   $R_{\rm
curv}\sim R_{\rm Schw}$).  The \gr\ emission from the accelerated protons is
thus expected in the  10~MeV -10~GeV energy region observable by {\it GLAST},
rather than in the TeV   region visible by  HESS.  
 
It is, in principle, not excluded that during short episodes of enhanced 
accretion the magnetic field can rise up to $10^3$~G, which would, in 
principle, allow  proton acceleration up to the energies $E_{max}\sim 
10^{20}$~eV. Thus, the energy of curvature emission   given by Eq.(\ref{curvp})
can  extend up to 10~TeV. Note, however,  that even in this case the observed
emission can not be related to the proton synchrotron  radiation which has an
intrinsic   self-regulated synchrotron cut-off at $\epsilon_{\rm synch}\le 
0.3$~TeV  \citep{synch_cut},  if the region  of proton acceleration is spatially
coincident with the region of synchrotron emission. A potential problem of 
assumption about transient  enhancement (by 4 orders of magnitude,   to produce
the necessary increase of  equipartition magnetic field, see Eq.(\ref{beq})) of
accretion   rate is that it should  result, in general, in a broad-band
flaring   activity of the nucleus of M~87,  which, however, is not observed. 

VHE \gr s are produced also in proton-proton ($pp$) collisions.
However, the interaction time of high-energy protons propagating through the 
low density medium ($n \leq 10^7$~cm$^{-3}$)  is quite large, 
\begin{equation} 
t_{pp}\simeq \frac{1}{\sigma_{pp}n_{max}}\simeq 10\left[\frac{10^7\mbox{ 
cm}^{-3}}{n_{max}}\right]\mbox{ yr}  
\end{equation} 
($\sigma_{pp}\sim 10^{-26}$~cm$^2$ is the proton-proton interaction 
cross-section). Even if $pp$ interactions would significantly contribute  to 
the overall VHE \gr\ emission, they cannot explain the fast day-scale \gr\ 
flux  variability of M~87.  Therefore the  observed variability should be 
referred  to fast changes in the concentration of multi-TeV protons in  the
source, i.e. due to adiabatic or  escape losses of protons,  on timescales  
comparable to $t_{\rm var}$.  The fast non-radiative losses versus slow rates 
of  \gr\ production at $pp$ collisions implies  very low efficiency of 
conversion of the energy of  parent protons to the VHE \gr s,  $\kappa=t_{\rm 
var}/t_{pp} \leq 3 \times 10^{-4}$. Thus  to explain the \gr\ luminosity $L_\gamma\sim  \rm
3\times  10^{40}$~erg/s, the proton acceleration power 
should exceed  $\kappa^{-1} L_\gamma\sim 10^{44}$~erg/s which is just about 
the luminosity  $L_{\rm Bondi}$  given by Eq. (\ref{bondi}), assuming a
conventional, 10\% or so, efficiency of conversion of the rest mass  energy of
accreting particles into  radiation.  (Here we ignore a possible  formation of
gamma-rays in a  relativistic outflow moving towards the observer which in
principle would  reduce by  an order of magnitude this requirement). 
 
The energy requirements to the proton acceleration power can be   somewhat
relaxed  if one invokes interactions of protons with the the surrounding
radiation fields.  Although TeV \gr s can be produced in a two-step process
which includes  Bethe-Heitler pair productions ($p \gamma \rightarrow p
e^+,e^-$) and  synchrotron  radiation of secondary electrons,   $p \gamma$
interactions become efficient when  they proceed through the
photomeson production channel.  In order to interact with the  photons of the
infrared source with average energy  $\epsilon_{\rm IR}\sim 10^{-2}$~eV,
protons should be accelerated to $E_{\rm p}  \sim   \left[200\mbox{
MeV}/\epsilon_{\rm IR}\right]m_p \sim 2 \times  10^{19}$~eV.   The number
density of IR photons in the compact infrared source is  
\begin{eqnarray} &&n_{\rm IR} = \frac{L_{\rm IR}}{4\pi R_{\rm IR}^2\epsilon_{\rm
IR}}\simeq\\ &&
7\times 10^{9}\left[\frac{L_{\rm IR}(\epsilon_{\rm IR})}{10^{41}\mbox{ erg/s}}\right]
\left[\frac{0.01\mbox{ eV}}{\epsilon_{\rm IR}}\right] \left[\frac{50 R_{\rm 
Schw}}{R_{\rm IR}}\right]^2 \mbox{cm}^{-3}   \nonumber
\end{eqnarray}  
For the average cross-section of the    photo-pion production cross-section,
$\sigma_{p\gamma}\sim 10^{-28}$~cm$^2$,   the  interaction time of protons with
infrared photons is   
\begin{equation} 
t_{p\gamma}=\frac{1}{\sigma_{p\gamma}n_{\rm IR}} \simeq 1.7
\left[\frac{10^{41}\mbox{ 
erg/s}}{L_{\rm IR}(0.01\mbox{eV})}\right] 
\left[\frac{R_{\rm IR}}{50 R_{\rm Schw}}\right]^2\mbox{ yr} 
\end{equation}  
If the infrared source is very compact, $R_{\rm IR}\sim R_{\rm Schw}/2$, and the  
the accretion rate is transiently increased 
by 2 orders of magnitude (to allow an 
order-of-magnitude increase in equipartition magnetic field and, as a consequence, proton 
acceleration to $E>10^{19}$~eV), the 
$p\gamma$ cooling time can be as short as the observed TeV variability  time scale.  
 
Note that the hypothesis of TeV \gr\ emission of M87   based on the  assumption of
$p \gamma$ interactions,    requires  a very compact IR source  with a size 
$R_{\rm IR}\sim R_{\rm Schw}$. This implies strong absorption of gamma-rays  
with fast    multiplication of electron-positron pairs  via Klein-Nishina
cascades.  Actually,   photon-photon pair production is an important element  
of any $p\gamma$ model; the observed spectrum   of TeV $\gamma$-rays cannot be
explained by   first generation of ultra-high  energy ($\geq 10^{15}$)  photons
from $\pi^0$ decays, and therefore requires    production of secondary
electrons which would provide   broad-band emission in the TeV energy band. On
the other hand,   the copious pair production may lead to    neutralization of
the large scale ($\geq  R_{\rm Schw}$)   electric field, and thus to
significant reduction of    the maximum  achievable energy of protons given  
by Eq.(\ref{emax}). Since   the rate of photomeson processes in the nucleus of
M87  is very sensitive to the energy of protons, namely, it requires   $E_{\rm
p} \geq 10^{18}$, the generation of   large amount of secondary electrons may
result   in a dramatic drop of the rate of photo-meson production.    A
non-negligible contribution to the secondary electrons  may come also from the
Bethe-Heitler pair production, especially    when the efficiency of photomeson
production  is suppressed  (the energy threshold of this process is two
orders   of magnitude smaller than the energy threshold of   the photomeson
production).   Whether this mechanism can explain the observed spectral  and
temporal characteristics  of TeV \gr\ emission   from M87, is a question which
needs detailed numerical  calculations.    In any case it is  clear that  $p
\gamma$ models  can provide adequate efficiency only in the case   of a very
compact IR source  with a size close to the Schwarzschild radius. 

\section{Gamma-ray emission from accelerated electrons.}  
\subsection{Order-of-magnitude estimates}  
\label{sec:order-of-mag} 
 
The tough requirements of acceleration of protons to   ultrahigh energies ($E
\geq 10^{18}$ eV),   as well as the relevant long cooling times   challenge any
interpretation of the   day-scale  variability of TeV \gr s in terms of    
interactions of high-energy protons. The models based on acceleration   of
electrons do not face such problems, and are  likely to be responsible for  
the observed TeV $\gamma$-ray  emission.  

The  main emission mechanisms by electrons   in the vicinity of the
supermassive BH are   synchrotron/curvature radiation and inverse  Compton (IC)
scattering.  Electrons can be accelerated to multi-TeV energies only if the
strength of the chaotic component of the magnetic    field, $B_{\rm rand}$, in
the acceleration region   is not too high. Assuming that electrons are  
accelerated at a rate   $dE/dt\sim \kappa eB_{\rm ord}$ ($\kappa \leq 1$ and
$B_{\rm ord}$   is the  ordered component of the magnetic field),    from the
balance of the  acceleration and  synchrotron  energy loss rates one finds   
\begin{eqnarray} 
\label{eq:max_e_synch} E_{\rm e} &\le& 
\frac{\kappa^{1/2}m_e^2B_{\rm ord}^{1/2}}{e^{3/2}B_{\rm rand}}\simeq  
\\&&4\times 10^{13}\left[\frac{B_{\rm ord}}{1 \mbox{ G}}\right]^{1/2}\left[  
\frac{B_{\rm rand}}{1  \mbox{ 
G}}\right]^{-1}  \kappa^{1/2} \mbox{ eV} \ . \nonumber 
\end{eqnarray}  
Thus, even in the case of maximum possible   acceleration rate ($\kappa=1$)
electrons  cannot  emit in the 10-100~TeV band unless   
\begin{equation}  
B_{\rm rand}\lsim  1\mbox{ G}  \ . 
\end{equation}    
In the ordered  field  the  energy dissipation  of electrons is   reduced to
curvature radiation loses. From the balance between  the curvature   loss rate
and the acceleration rate, assuming that  the typical curvature radius $R_{\rm
curv}$ of magnetic field  is   comparable to   the gravitational radius, one
finds 
\beqa  
\label{eq:max_e_curv} 
E_e&=&\left[{3m_e^4 R_{\rm curv}^2 \kappa B_{\rm ord} \over 2e}\right]^{1/4}\simeq \\ 
&&4\times 10^{15} 
\left[\frac{B_{\rm ord}}{1 \mbox{ G}}\right]^{1/4}\left[ 
\frac{R_{\rm curv}}{R_{\rm Schw}}\right]^{1/2}\kappa^{1/4}\mbox{ eV} \ .  
\nonumber  
\eeqa  
Thus if the energy losses of electrons   are dominated by curvature
radiation,   the maximum energy  of accelerated electrons   only weakly depends
on the strength of  the magnetic field. 
 
The IC loss rate is determined by the energy density of infrared  
radiation in the nucleus, $U_{ph}=L_{\rm IR}/(4\pi R_{\rm IR}^2c)$.  The
condition of the balance between IC loss rate and the electron   acceleration rate
gives 
\beqa 
\label{eq:max_e_ic} 
E_{\rm e}&\simeq& \frac{3^{3/4}m_e^2B_{\rm ord}^{1/2}R_{\rm IR}}{2^{5/4}e^{3/2} 
L_{\rm IR}^{1/2}}\simeq 
\\ &&1\times 10^{15}\left[\frac{B_{\rm ord}}{1\mbox{ G}}\right]^{1/2} 
\left[\frac{R_{\rm IR}}{50 R_{\rm Schw}}\right]\kappa^{1/2} 
\mbox{ eV}  \ . \nonumber 
\eeqa 
Note that this  estimate is obtained  assuming that the IC scattering   takes
place in the Thompson regime. However, highest energy  electrons upscatter the 
infrared/optical radiation   in the Klein-Nishina regime in which the
efficiency of the IC scattering is reduced. A proper account of the  decrease of
the IC loss efficiency  would result in
higher electron energies exceeding the estimate of Eq.
(\ref{eq:max_e_ic}).

\subsection{Electron acceleration in the vacuum gaps of BH magnetosphere} 
 
So far  we did not specify the particular mechanism of particle  acceleration. 
In principle, several  mechanisms can be  responsible for the electron
acceleration,   but an obvious requirement which follows from the above 
estimates is that the "efficiency" parameter $\kappa$ for the  acceleration
rate should be close to one, otherwise electrons would not reach  the $\sim
100$~TeV energies, as it follows from Eqs. 
(\ref{eq:max_e_synch}),(\ref{eq:max_e_curv}),(\ref{eq:max_e_ic}).   Also, the
irregular component of the magnetic field should not exceed  1~G. 

Large scale ordered electric fields, induced by rotation of black hole, 
are known to be 
responsible for particle acceleration and  high-energy radiation 
in  pulsars. A similar mechanism of generation of large electric fields can  be
realised  in the vicinity of a rotating  BH placed in an external magnetic
field    \citep{wald,bicak}. In the case  of pulsars, it is known that the
force-free magnetosphere possesses so-called  "vacuum gaps" in which the
rotation-induced electric field is not neutralized by  redistribution of
charges. The vacuum gaps work  as powerful particle  accelerators and sources
of pulsed high-energy \gr\ emission. Vacuum gaps with strong   rotation-induced
electric field can be present also in the vicinity of a rotating
 black hole
\citep{blandford77,beskin92}. Below we explore whether the observed VHE \gr\  emission from
\m87\ can be explained by the emission from the vacuum gaps   formed in the
magnetosphere of the supermassive black hole in M87. 

\subsubsection{The magnetosphere of rotation-powered black hole} 
\label{sec:magnetosphere} 
 
Throughout the magnetosphere  the component of electric field directed
along the magnetic field lines  is neutralized by the charge redistribution, so
that a  force-free condition $\vec B\bot\vec E$ is satisfied. 
The characteristic charge density needed to neutralize the parallel component of
electric field 
in the  magnetosphere  of a BH rotating with an angular velocity  $\vec\Omega$
placed in an external magnetic field $\vec B$ is the so-called 
"Goldreich-Julian" density
\citep{goldreich} 
\begin{equation} 
\label{GJ} 
n_q\simeq \frac{\vec\Omega\cdot\vec B_{\rm ord}}{2\pi e} 
\simeq\frac{aB_{\rm ord}}{(GM)^2}  
\end{equation} 
($a=\Omega(GM)^2$, the BH rotation moment  per unit mass, $0<a<GM$, is a
commonly used parameter of the Kerr  metric describing the space-time of
rotating black hole; see  Appendix).  

In general, the charge distribution in the magnetosphere  is not  static
-- additional free charges should be continuously supplied  throughout the
magnetosphere to compensate for the charge loss due to the magnetohydrodynamical
outflow. The inefficiency of charge supply can lead to the formation of "gaps"
in the magnetosphere in which the parallel component of electric field is not
zero and conditions for particle acceleration exist.  

In the case of pulsars, there are several  potential ways to supply charged
particles to the magnetosphere. First  of all, the charge can be extracted
directly from the   surface of the neutron star.   Electrons and positrons can
be generated also due to   pair production in very strong magnetic field.  
Finally, electron-positron pairs can be created at interactions of
$\gamma$-rays with low energy photons.   

Apart from the extraction of free charges from the surface of the compact 
object, the same mechanisms can in principle, be responsible for the charge 
supply to the magnetosphere in the case of black holes. However, in the
particular case of the black hole in M87, the pair production of \gr s in the  
magnetic field, (a mechanism, assumed e.g. in the \citet{blandford77} scenario) is not efficient because (1) the magnetic field cannot  
significantly  exceed 10 G and (2) the energy of $\gamma$-rays   emitted by
accelerated particles cannot exceed 100 TeV. On the other hand,  the efficiency
of the charge supply via the pair production by \gr s on the soft infrared
background depends on the compactness of the infrared  source ($\propto L_{\rm
IR}/R$). This process can be efficient only if \gr s with energies above 10~TeV
are present in the compact source. 

Since the 10~TeV \gr s have to be produced by particles accelerated to 
energies  above 10~TeV, the gap(s), in which electric field component along the
magnetic field lines is not neutralized by the charge redistribution,  should 
be present in the magnetosphere. In a self-consistent scenario
the height of the gap(s) is limited by the condition that   \gr s emitted by
the  accelerated particles do not produce $e^+e^-$ pairs within the gap. 

In order to estimate whether particle acceleration and high-energy  emission
from the vacuum gaps can be responsible for the observed VHE  luminosity of
\m87\ one has to estimate the total acceleration   power output in the gap.  In
spite of the fact that   the potential drop in  the gap can be enough to
accelerate charge   particles to energies as high  as $10^{18}$~eV, strong
radiative losses limit the     maximum energy of electrons to $10^3$~TeV. This
means that the  propagation of electrons through the gap proceeds in a 
"loss-saturated" regime: all the work done by the gap's electric field  is
dissipated through the synchrotron/curvature and/or IC  radiation.  The
rotation induced electric field near the   BH horizon has a strength (see
Appendix)   ${\cal E}\sim \left[a/GM\right]B_{\rm ord}$. For each  electrons
propagating in the gap,  the energy loss rate   is estimated as $dE/dt\simeq
e{\cal E}\sim  e\left[a/(GM)\right]B_{\rm ord}$.\footnote{This implies   that
the acceleration  efficiency  $\kappa$ is $\kappa\simeq  \left[a/(GM)\right]$.
For the extreme   rotating black hole with   $a=GM$, the acceleration reaches
the maximum possible rate with    $\kappa\simeq 1$.}  The density of electrons
in the gap is limited by the  Goldreich-Julian density, given by Eq. (\ref{GJ}).  If the size
of the infrared  source is large enough, so that the gap height is not limited
by   pair production, the size of the gap is estimated to be about $H\sim 
R_{\rm Schw}$.  Taking into account that the volume of the gap is  roughly
$R_{\rm Schw}^2H$, the total number of electrons in the gap  can be estimated
as $R_{\rm Schw}^2Hn_q$. Then the   total power output of the gap is 
\begin{eqnarray} 
\label{power} 
P&\simeq &n_qHR_{\rm Schw}^2(dE/dt) \\&\sim&  
5\times 10^{41}\frac{4a^2}{R_{\rm Schw}^2}\left[\frac{M}{3\times 
10^9M_\odot}\right]^2\left[\frac{H}{R_{\rm Schw}}\right]\left[\frac{B_{\rm ord}}{10\mbox{ G}}\right]^2 
\mbox{ erg/s}\nonumber 
\end{eqnarray} 
 
Thus, if the angular momentum of the black hole  
is large enough, the nonthermal power of the  
vacuum gap can be as large as the observed  
TeV gamma-ray luminosity.

\subsubsection{Numerical modelling of acceleration and radiation of  
electrons in the gap of the black hole magnetosphere} 
 
Location of the gaps in the BH magnetosphere depends on the structure  of both
the accretion flow and the magnetic field near the event   horizon. In this
regard, it should be noted that even after  four decades of intensive
theoretical study of physics of   pulsar magnetospheres, the details of the
geometry of   vacuum gaps remain uncertain. Nevertheless,   the basic
properties of particle accelerators operating in   the vacuum gaps can be
understood with a reasonable  accuracy and confidence. 

We have developed a numerical code which allows,  for the given geometry of the
gap and configuration of  the magnetic field, a quantitative   study of energy
distributions of electrons accelerated in   the vacuum gap and associated
electromagnetic radiation.    For demonstration of the  importance of the VHE
\gr\ emission from the vacuum gaps, in this   paper we have chosen a simple
geometry of the gap, namely  we assumed that the gap occupies a  spherical
layer above the BH horizon and has the height of about the  size of the event
horizon. The geometry of electromagnetic field is  assumed to be given by the
solution of Maxwellian equations in Kerr  metric, which corresponds to an
asymptotically constant magnetic  field inclined at an angle $\chi$ with
respect to the rotation axis of  the black hole \citep{bicak}. The analytical
solution of    Maxwellian  equations are given in Appendix.   The initial
locations of electrons are assumed to be  homogeneously distributed either
throughout the gap or over the  boundary of the gap. The initial energies of
electrons are assumed   to be equal to the rest energy.   Trajectories have
been  numerically integrated taking into account  effects of  General
Relativity and energy losses  of electrons due to synchrotron/curvature
radiation and  inverse Compton scattering   (see Appendix for technical
details). The spectra and angular  distributions of the synchrotron and
curvature radiation are  calculated by tracing the photon trajectories through
the Kerr  space-time metric from the emission point to infinity. 

\begin{figure*} \begin{center} 
\includegraphics[width=\linewidth]{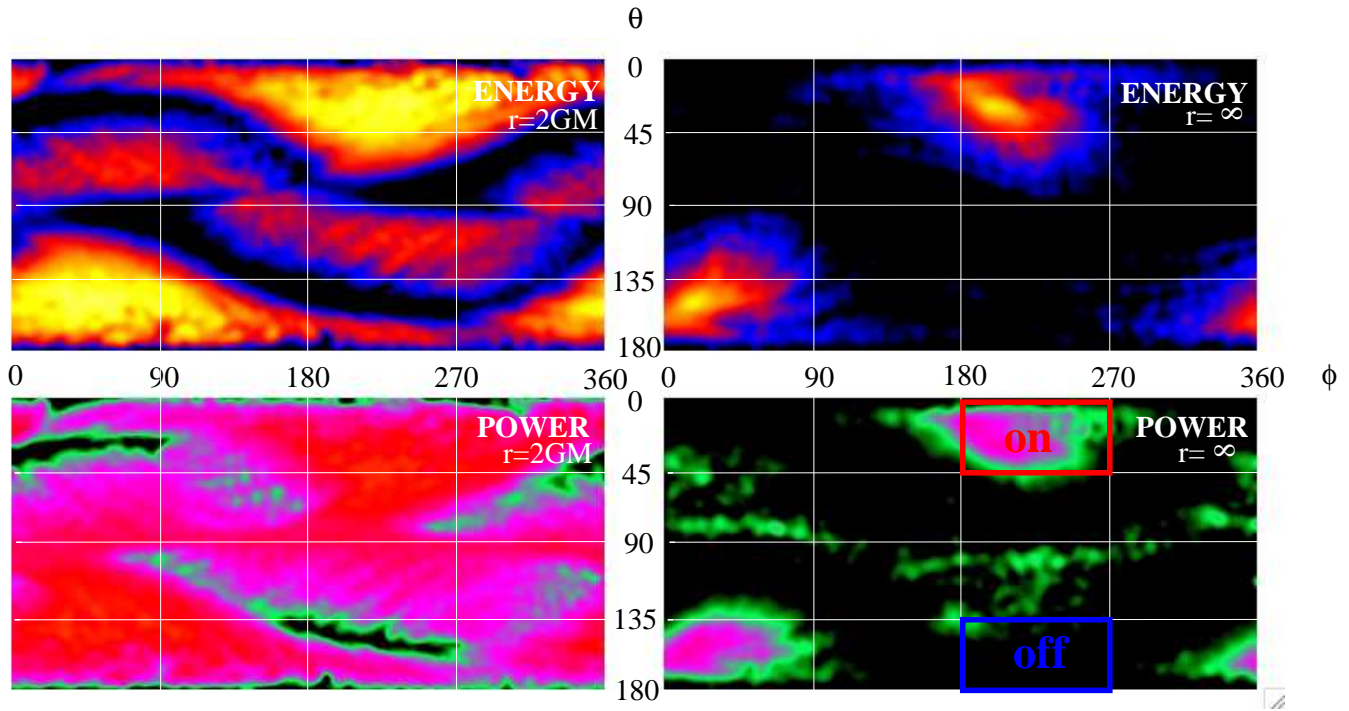} 
\caption{Angular  distributions of synchrotron/curvature photon energies  (top)
and \gr production rate   (bottom) for  emission from a spherical vacuum gap 
close to an  extreme rotating BH placed in an magnetic field $B=1$~G inclined
at an angle  $\chi=20^\circ$ with respect to the BH rotation axis. Two left
panels show  the angular distributions as seen in ZAMO frame    close to the
black hole horizon. Two right panels show the angular distributions   after
photon tracing to infinity. The scales are logarithmic and cover two decades
from  maximum (yellow on the top panels, red on the bottom panels) scale,
$max$, down  to the $0.01max$ (black). Red and blue boxes on the bottom right 
panel show the regions used to extract spectra seen from "on hot spot" and "off
hot spot" directions as shown in Fig. \ref{fig:th20_on_off}. }
\label{fig:map_th20} 
\end{center}  
\end{figure*}

As an example of numerical modelling, in Fig. \ref{fig:map_th20} we show  
distributions of average energies of electrons propagating   in the spherical
layer occupied by the gap. The magnetic field is   assumed to be inclined at an
angle  $\chi=20^\circ$ with respect to the  rotation axis. The two left panels
show angular distributions of the average energy of individual electrons
and the   power of synchrotron/curvature emission, as  measured in the Zero
Angular Momentum Observer (ZAMO) frame \citep{bardeen},   close to the black
hole  horizon (see Appendix for details). One can see that the maximum 
energies of photons are achieved in two oppositely situated hot-spots
determined  by the direction of magnetic field. At the same time, the total
power of radiation  does not strongly depend on the latitude and longitude
coordinates.  

The "dark strips" (one along the equatorial plane and two snake-like dark
strips  above and below the equatorial plane) are clearly recognizable  in the
left panels of  Fig. \ref{fig:map_th20}. The drop of   energies  is explained
by the specific configuration of electromagnetic field in  these regions.
Namely, the dark strips surround the so-called  "force-free"  surfaces at which
the rotation-induced electric field $\vec E$  is orthogonal  to the  magnetic
field $\vec B_{\rm ord}$. 

The right panel of Fig. \ref{fig:map_th20} shows the angular distribution of 
average photon energies and emissivity after the photon tracing to infinity through the Kerr 
space-time metric. One can see that polar "hot spots" become more pronounced, mostly 
because the  photons emitted from equatorial regions (from the 
ergosphere) have larger  redshifts and a significant fraction of these photons 
is just absorbed by the black hole. 
 
\begin{figure} 
\begin{center} 
\includegraphics[width=\linewidth]{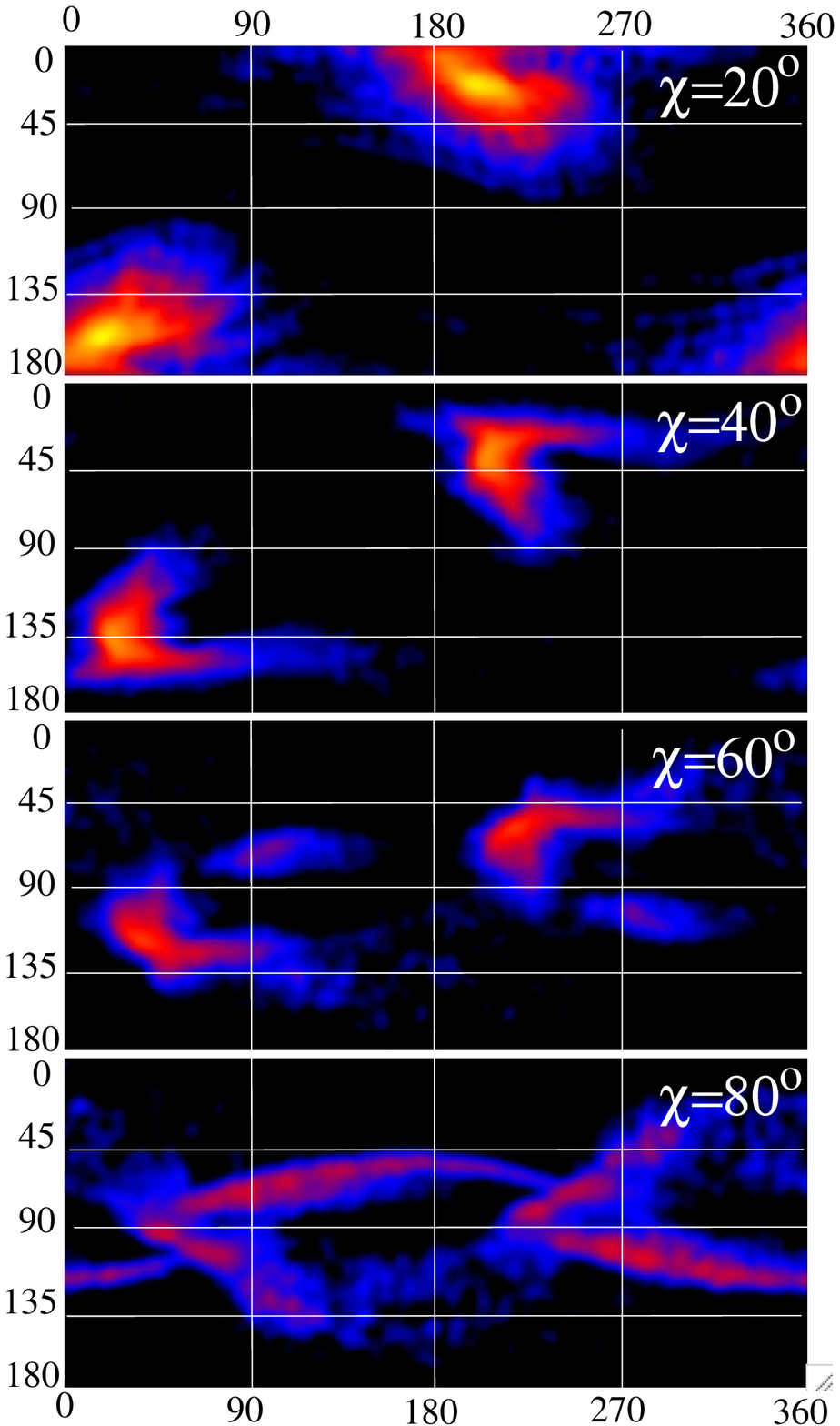} 
\caption{Evolution of the shape of the polar hot spots with an increase 
of the inclination angle of magnetic field. The figures show 
the angular distribution of the energies of photons of synchrotron/curvature
radiation traced to infinity. The color scale and  
parameters of numerical simulations are the same as in the top right panel of 
Fig. \ref{fig:map_th20}: maximum yellow corresponds to photon energies 10~GeV, 
minimum (black) to photon energies below 0.1~GeV.} 
\label{fig:map_th20_40_60_80} 
\end{center} 
\end{figure}

Fig. \ref{fig:map_th20_40_60_80} shows evolution of the shape of the polar  hot
spots with an increase of the inclination angle of the magnetic field. One can  see
that the shape of the hot spots becomes wider and irregular. Also,   with an
increase of the  inclination of the magnetic field   the average  photon energy
decreases, which is explained by the fact that   the acceleration of electrons
is most efficient when   the electric field is aligned with the magnetic 
field. With the increase of the magnetic field inclination angle $\chi$, the  
regions of aligned electric and magnetic field (situated close to the rotation 
axis of the BH in the case $\chi=0^\circ$)  disappear. 

\begin{figure}  
\begin{center} 
\includegraphics[width=\linewidth]{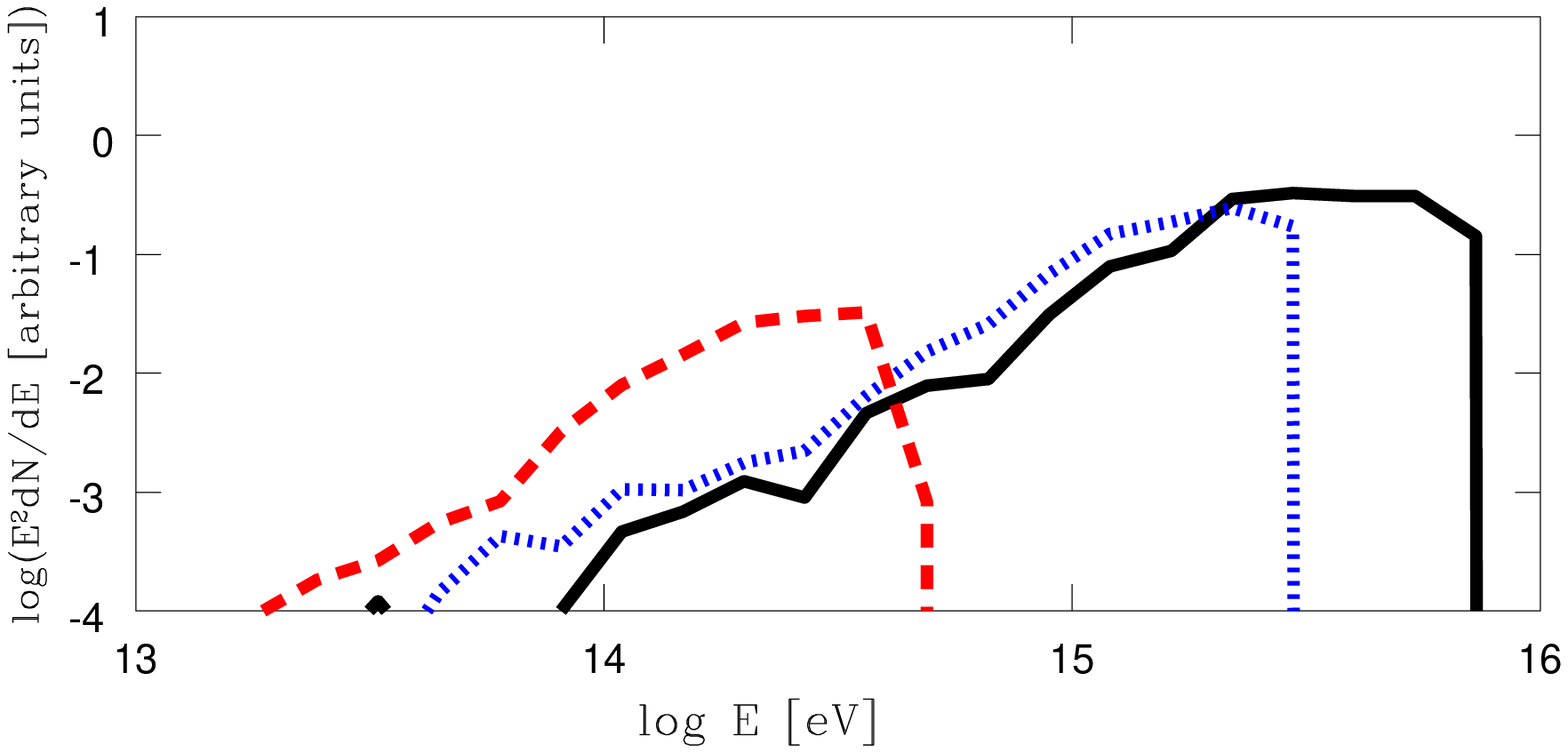}  
\caption{Spectrum of electrons accelerated in the vacuum gap above the horizon of a 
maximally rotating BH of a mass $M=3\times 10^9M_\odot$ placed in 
an ordered magnetic field $B=1$~G inclined at $\theta=60^\circ$  
with respect to 
the BH rotation axis. Black solid line: the random magnetic field is 
0.1\% of the ordered one; blue dotted line: the random magnetic field is 1\%; 
red dashed line: random magnetic field is 10\%.} 
\label{fig:electrons}  
\end{center}  
\end{figure}

In Fig. \ref{fig:electrons} a typical spectrum of electrons accelerated  in the
spherical vacuum gap close to the BH horizon is shown. It is assumed that   an
extreme rotating BH ($a=GM$) is placed in  $1$~G magnetic field inclined at an
angle of $60^\circ$ with respect to the  rotation axis. The size of the
infrared emission region is assumed $R_{\rm  IR}=10R_{\rm Schw}$. The three
energy spectra shown in Fig. \ref{fig:electrons}  correspond to different
strengths of the   random component of magnetic field. If the random magnetic
field is smaller  than $10^{-3}$ fraction of the ordered field,    electrons
propagating in the gap reach  energies up to $\sim 10^{16}$~eV. With an
increase of the   random component of magnetic field, the synchrotron losses
start to dominate  which leads to reduction  of the  maximum electron energies
energies,   in a good agreement with the qualitative estimates of Section 
\ref{sec:order-of-mag}. 

\subsection{Direct synchrotron/curvature and IC radiation from the 
acceleration process} 
 
The radiative losses through both synchrotron/curvature and IC channels are
released  in the form  of high energy $\gamma$-rays.  The energy of Compton
upscattered photons (in Thompson regime) is 
\begin{equation}  
\epsilon_{\rm IC}=0.4\left[\frac{\epsilon_{\rm IR}}{10^{-2}\mbox{eV}}\right] 
\left[\frac{E_e}{1 \mbox{ TeV}}\right]^2\mbox{ TeV}  
\end{equation} 
The IC scattering of $E_e\gsim 10$~TeV electrons   on IR photons proceeds in
the Klein-Nishina regime, thus    $\epsilon_{\rm IC}\simeq E_e$. The  
curvature radiation peaks at significantly lower  energies,    
\begin{equation} 
\label{curv_e} 
\epsilon_{\rm curv} ={3 E_e^3\over 2 m_e^3R_{\rm curv}} 
\simeq 0.2\left[\frac{E_e}{10^{15}\mbox{ 
eV}}\right]^3 \left[\frac{R_{\rm Schw}}{R_{\rm curv}}\right] \mbox{ GeV} \ . 
\end{equation}  
 
Since electron acceleration in the gap proceeds in the "loss saturated" regime,
the calculations of the spectral and angular distributions of
radiation   accompaning the acceleration process,  requires  
``self-consistent''   approach  in which the spectrum of radiation is
calculated simultaneously with  the  spectrum of  parent  electrons. The
algorithm of  self-consistent calculations used in this work is briefly  
described in the Appendix.  

\begin{figure}  
\begin{center} 
\includegraphics[width=\linewidth]{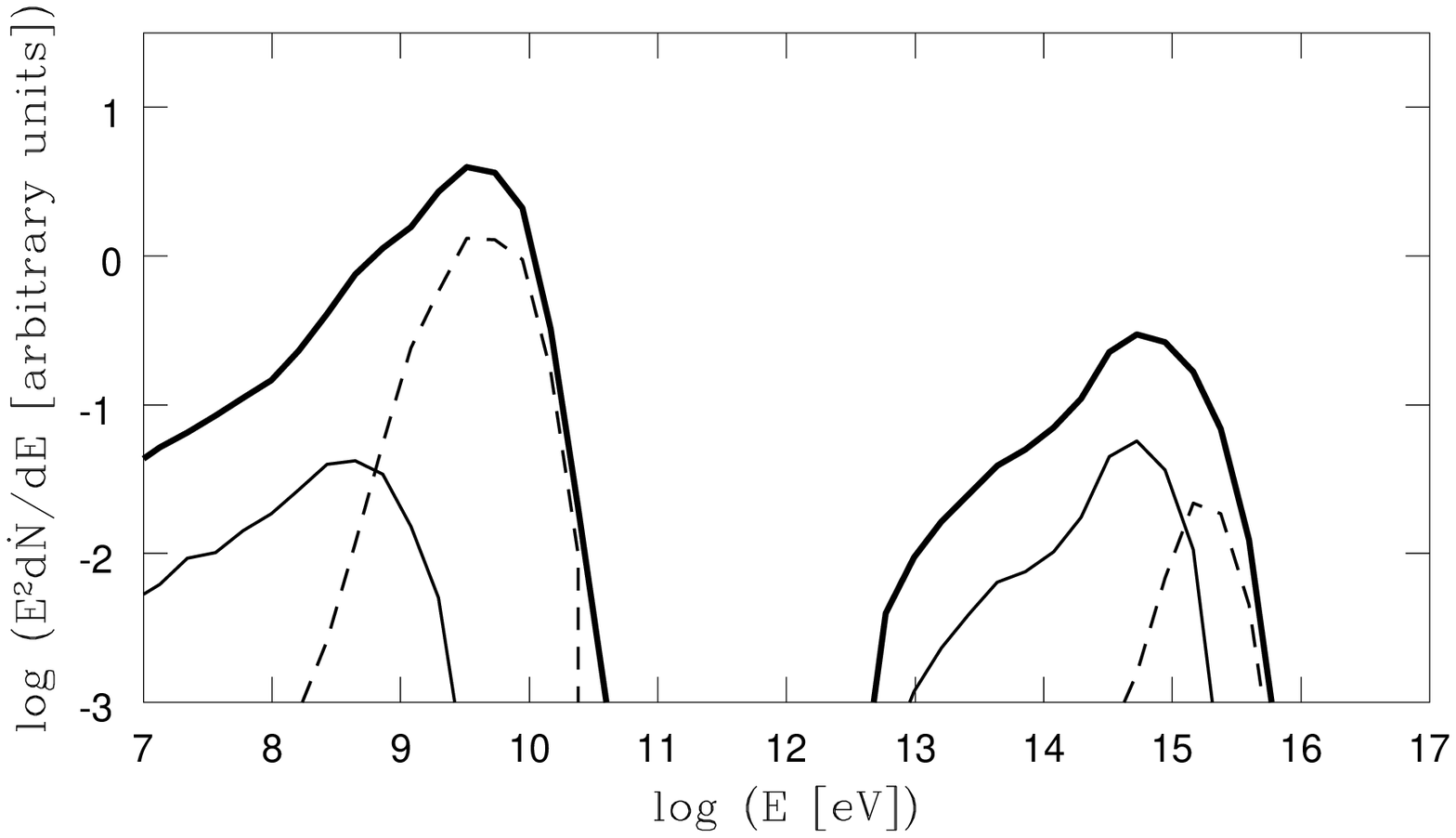}  
\caption{The production rate  of first generation gamma-rays  
emitted by  the spherical vacuum gap. The 
physical parameters are the same as in Fig. \ref{fig:map_th20}. Thick solid 
line: the total spectrum integrated over all directions. Dashed line: the  
spectrum integrated over the direction around the "hot spot" (the box marked 
"on" in Fig. \ref{fig:map_th20}). Thin solid line: the spectrum collected from 
the box marked "off" in Fig. \ref{fig:map_th20}.} 
\label{fig:th20_on_off}  
\end{center}  
\end{figure} 
 
\begin{figure} 
\begin{center} 
\includegraphics[width=\linewidth]{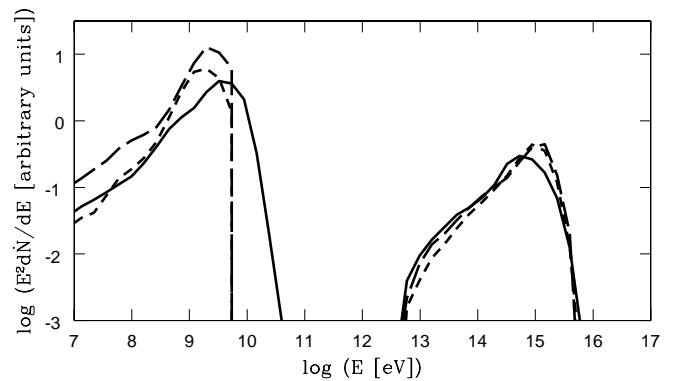} 
\caption{The production rates of gamma-rays calculated  for different values of
the inclination angle of  magnetic field.  Solid line: $\chi=20^\circ$;  dashed
line: $\chi=60^\circ$;  long-dashed line: $\chi=90^\circ$. Physical parameters
are  same as in Fig.   \ref{fig:map_th20}.  } 
\label{fig:direct} 
\end{center} 
\end{figure} 
 
Some results of self-consistent calculations of the    $\gamma$-ray production
spectra as  functions of   the viewing angle and the inclination of the
magnetic field are shown   in Figs.  \ref{fig:th20_on_off} and
\ref{fig:direct}, respectively.   Fig.\ref{fig:th20_on_off} demonstrates  the
difference   of production spectra of $\gamma$-rays emitted along  the
direction of magnetic field (the region marked "on" in Fig. 
\ref{fig:map_th20}) and away from this direction (the region marked "off" in
Fig.  \ref{fig:map_th20}). Fig. \ref{fig:direct} demonstrates   the dependence
of the \gr\ production spectra on   the inclination angle of magnetic field. In
both figures   the low-energy (MeV-GeV)  peak is due to the
synchrotron/curvature   radiation, while the high energy (TeV-PeV) peak is
formed due to the   IC scattering in the Klein-Nishina regime.  

\subsection{Isotropic TeV emission from secondary pair-produced  
electrons} 

\begin{figure} 
\begin{center} 
\includegraphics[width=\linewidth]{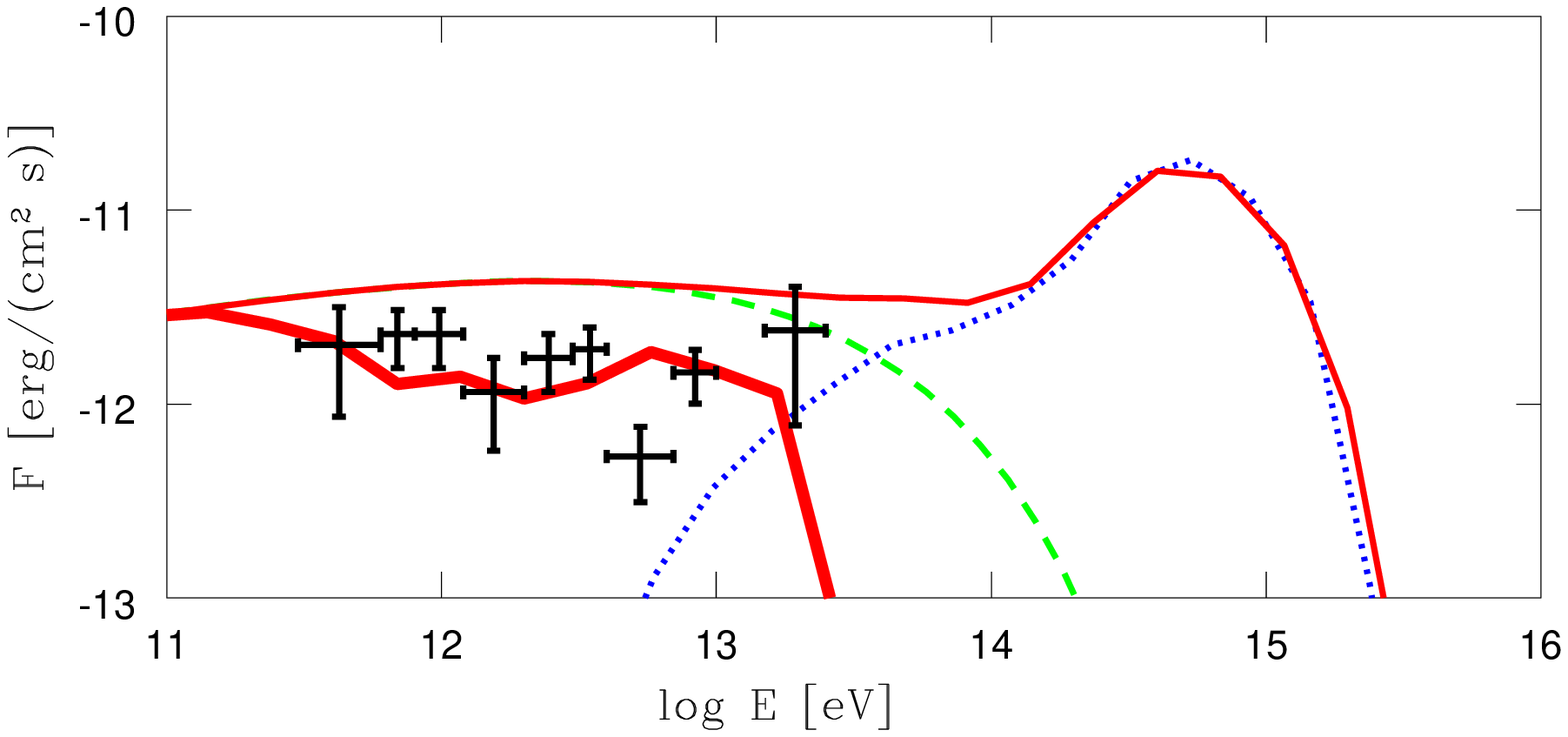} 
\caption{Secondary emission from high-energy electrons injected into the compact 
infrared source in the nucleus of \m87\ via the photon-photon pair production.  
Thick black solid line: omnidirectional spectrum of primary emission from accelerated 
particles; thin blue solid line: the primary spectrum only from the "off" 
direction (same as in Fig. \ref{fig:th20_on_off}); black and blue thin  
dashed lines show the spectra attenuated by the pair production in the infrared 
source. Red dotted line shows the contribution of secondary cascade (isotropic) 
emission. 
} 
\label{fig:cascade} 
\end{center} 
\end{figure} 
 
The spectral energy distributions shown in Figs.  \ref{fig:th20_on_off} and
\ref{fig:direct} correspond to the production rates of the first generation  
$\gamma$-rays. They have essentially anisotropic distribution, thus the  
calculations of fluxes detected by an observer contain large uncertainties,
mainly   because of the poor knowledge of the source geometry. However, due to
the   internal and external absorption of $\geq 10$ TeV $\gamma$-rays, the
observer   detects only a tiny fraction of the first generation $\gamma$-rays.
While  interactions with external photon fields lead to   real attenuation of
the \gr flux, the internal absorption is essentially   recovered due to
radiation of the pair produced electrons of   second and further generations. 
Interestingly, the   development of an electromagnetic cascade in  radiation
field of the infrared source may lead to ``isotropisation'' of  the \gr\
source.   Indeed,   the absorption of first generation $\gamma$-rays   leads to
deposition of $e^+e^-$ pairs throughout the infrared  source volume, $R_{\rm
IR}$. If the latter is significantly larger   than the volume corresponding to
the vacuum gap (i.e. $R_{\rm IR} \gg R_{\rm Schw}$),   the magnetic field in
the IR source   can be dominated by the irregular component which would
effectively   isotropise the directions of secondary electrons.
Correspondingly,   the secondary radiation  from $e^+e^-$ pairs will be 
emitted  isotropically. For  any reasonable magnetic field, the synchrotron  
radiation of secondary electrons is produced at energies   significantly below
1 TeV. Therefore for explanation of the observed   TeV gamma-radiation one
should assume that the energy losses of   electrons are dominated by IC
scattering,   i.e. the magnetic field in the   infrared source should be
significantly less than   $B=(L_{\rm IR}/2 R_{\rm IR}^2 )^{1/2} \sim 0.1\ \rm
G$.  If so,   the absorption  of first generation gamma-rays will trigger an  
electromagnetic (Klein-Nishina) cascade.  

In Fig. \ref{fig:cascade} we show the resulting spectrum of   gamma-radiation
expected  from the internal absorption of first   generation gamma-rays. It
consists of the isotropic component  associated with the  cascade in the
infrared source (green dashed curve)   and the primary anisotropic component
whose intensity is uncertain since it strongly depends on the orientation of
the observer with respect to the magnetic field direction. In  Fig.
\ref{fig:cascade} the thin solid red line   corresponds to the sum of these two
components, while the thick red solid line 
shows the result of absorption of the summary spectrum  
  in the  infrared source, in the
elliptical galaxy M87 and in the intergalactic medium  (see thick solid curve
in Fig. 1). The curve is normalized to the observed   flux of $\gamma$-rays at
0.5 TeV. The comparison of the calculated \gr\ spectrum    shows quite a good 
agreement with the HESS measurements   up to  $E \sim 10-20$ TeV. One should
note, however, that the agreement with the observations should not be
over-emphasized, since we consider a "toy model" aimed to demonstrate the
importance of TeV emission from the vacuum gaps in the magnetosphere. 

\begin{figure} 
\begin{center} 
\includegraphics[width=\linewidth]{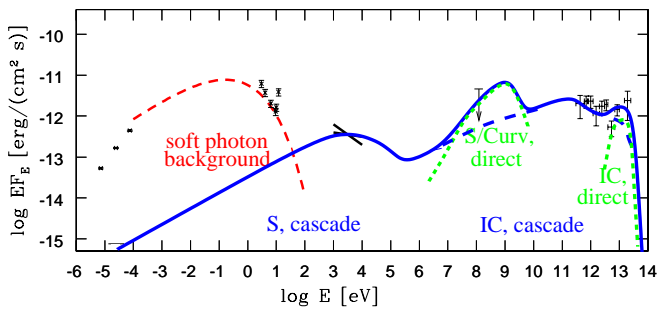} 
\caption{Spectral energy distribution of emission from the nucleus of \m87.
Dashed  
blue line: isotropic synchrotron and IC emission from secondary 
electron 
positron pairs injected via the pair production. Short-dashed green line:
direct synchrotron/curvature and IC emission from electrons accelerated in the
vacuum gap (strongly anisotropic, depends on the geometry of the vacuum gap). 
Solid blue line shows the total emission spectrum, which is the sum of the direct
and cascade contributions.
Dashed  red line shows the model spectrum of soft photon background used for the 
calculation of IC scattering. This radiation component can come 
from a larger region and is not necessarily related to the particle acceleration
in the vacuum gap. 
} 
\label{fig:SED} 
\end{center} 
\end{figure}

Finally in Fig. \ref{fig:SED} we show the broad-band spectral energy
distribution (SED)  of the resulting radiation  and compare the model curve
with observed   fluxes of the nucleus of M87 at infrared, X-ray and TeV
gamma-rays.  Two broad peaks in the SED correspond to synchrotron radiation and
inverse Compton   scattering of secondary (cascade) electrons in the infrared
source.   The condition that the synchrotron emission from the infrared source
should not exceed the observed flux in the X-ray band imposes an upper limit 
on the random magnetic field strength $B<0.1(R_{IR}/R_{\rm Schw})^{-1}$~G. 
The existing upper limit on the M87 flux in the EGRET energy band
\citep{sreekumar94} imposes a restriction on the direct synchrotron/curvature
emission from gap emitted in the direction of observer. 

Because of uncertainties of model parameters as well as  
the strong variability of radiation, we do not 
attempt to make a detailed fit to the broad band spectrum of the source, 
especially taking into account that the measurements at different energy bands  
correspond different epochs.  On the other hand, the future simultaneous 
studies  of temporal and spectral properties of the broad-band  
emission can provide meaningful tests of the proposed model and  
significantly reduce the relevant parameter space.

\section{Summary and Conclusions.} 
 
We have studied the mechanisms of production of variable   TeV \gr\ emission
from vicinity of the supermassive BH in the nucleus of  \m87. Moderate
accretion rate onto the black hole, inferred  from the {\it Chandra}
observations, limits the magnetic field strength close to the  black hole
horizon - it  cannot significantly   exceed $B \leq 10$~G. This limits the
maximum energy  attainable by protons,   $E \leq 10^{18}$~eV. None of the known
mechanisms of \gr\ emission by   protons can satisfactorily explain the
observed temporal and spectral  characteristics of the observed TeV \gr\
emission.   On the other hand, severe radiative losses of electrons in a
compact   region close to the black hole significantly constrains the range
of   possible acceleration scenarios. In particular the random component   of
the magnetic field cannot exceed 1 G. Thus,  the acceleration takes place, 
most likely, in a region where the regular magnetic   field significantly
exceeds the random component of the field. Even so,  the unavoidable energy
losses due to the curvature radiation and   inverse Compton scattering require
an extremely effective mechanism   of particle acceleration   with a rate close
to the maximum   (theoretically  possible) acceleration rate.  

In this paper we show that the observed TeV gamma-ray emission   from M87 can
be explained by electrons accelerated   in strong rotation induced  electric
fields in the vacuum gaps in black hole magnetosphere. Generally, this model
has many similarities with models of particle acceleration in pulsar
magnetospheres.  Our detailed modelling shows that the 
gamma-radiation from the central engine of M~87 consists of
both   first generation photons emitted by particles accelerated in the gap
(severely attenuated due to interactions
with the internal   and external radiation fields) and  second and further
generation   (cascade) photons. If the first component   dominates above 10
TeV, the cascade \gr s contribute mainly to  the $\leq 10$~TeV  energy domain.
The electron acceleration and \gr\ production in a very compact   region close
to the event horizon of the black hole naturally explains the observed  
variability of TeV \gr\ emission from M87.   

\section{Acknowledgement.}

We would like to thank V.Beskin for the clarifying comments on he manuscript. 
\newpage 
\appendix 
\section{Details of the numerical modelling of particle acceleration in 
the vacuum gaps in magnetospheres of rotating black hole.} 
 
A self-consistent modelling of acceleration of electrons in the direct vicinity 
of event horizon of a BH requires (a) a full account of the effects of 
General Relativity and (b) a full account of the radiation reaction on particle 
motion, since electrons propagate most of the time in  the  
"loss saturated" regime when the acceleration force is balanced by the radiation 
reaction force. Below we give some details of the modelling 
of trajectories electrons and photons in the vicinity of the black hole.

\subsection{The Kerr space-time.} 
 
A rotating  BH is described by two  
parameters: its mass $M$ and the angular momentum per unit mass $a\le GM$.  
The geometry of space-time in the vicinity of horizon is described by the  
Kerr metric  
\beq 
\label{kerr} 
ds^2=-\alpha^2dt^2+g_{ik}\l(dx^i+\beta^idt\r)\l(dx^k+\beta^kdt\r) 
\eeq 
$$ 
\alpha=\frac{\rho\sqrt{\Delta}}{\Sigma};\ \ g_{rr}=\frac{\rho^2}{\Delta};\ \  
g_{\theta\theta}=\rho^2;\ \ g_{\phi\phi}=\frac{\Sigma^2\sin^2\theta}{\rho^2}; 
$$ 
$$ 
\beta_\phi= 
-\frac{2aGMr}{\Sigma^2}; \ \ \Delta=r^2+a^2-2GMr; 
$$ 
\beq 
\Sigma^2=(r^2+a^2)^2-a^2\Delta\sin^2\theta;\ \  
\rho^2=r^2+a^2\cos^2\theta 
\eeq 
The horizon is situated at $r_H=GM+\sqrt{(GM)^2-a^2}$.  
 
To understand the acceleration and energy losses of charged particles propagating 
close to the BH horizon, it is convenient to use an orthonormal (non-coordinate) frame  
\beqa 
\label{zamo} 
e_{\hat 0}=\frac{\Sigma}{\rho\sqrt{\Delta}}\frac{\partial}{\partial t}+ 
\frac{2GMar}{\Sigma\rho\sqrt{\Delta}}\frac{\partial}{\partial \phi};\ \  
e_{\hat r}=\frac{\sqrt{\Delta}}{\rho}\frac{\partial}{\partial r};\ \  e_{\hat\theta}=\frac{1}{\rho}\frac{\partial}{\partial\theta};\ \  
e_{\hat\phi}=\frac{\rho}{\Sigma\sin\theta}\frac{\partial}{\partial\phi}. 
\eeqa 
carried by the so-called "zero angular momentum" observers (ZAMO) \citep{bardeen}. 
The corresponding covariant basis vectors  $e^{\hat i}$ are given by 
\beqa 
\label{zamo1} 
e^{\hat 0}=\frac{\rho\sqrt{\Delta}}{\Sigma}dt;\ \  
e^{\hat r}=\frac{\rho}{\sqrt{\Delta}}dr;\ \  
e^{\hat \theta}=\rho d\theta,\ \  
e^{\hat \phi}=\frac{\Sigma\sin\theta}{\rho}d\phi-\frac{2GMar\sin\theta}{\rho\Sigma}dt. 
\eeqa 
 
\subsection{The electromagnetic field.} 

In the reference frame (\ref{zamo})  the magnetic field inclined at angle $\chi$ with  
respect to the BH rotation axis is given by \citep{bicak}  
\beqa 
\label{B} 
B^{\hat r}=\frac{1}{\Sigma\rho^4\sin\theta}&&\left\{ 
B_{\|}\sin\theta\cos\theta\left[\Delta\rho^4+2GMr(r^4-a^4)\right]+ 
\right.\nn\\&&\left. 
B_\bot 
\left[r\cos\psi-a\sin\psi\right] 
\left[\rho^4(r\sin^2\theta+GM\cos 2\theta)- 
\right.\right.\nn\\&&\left.\left. 
GM(r^2+a^2)(r^2\cos 2\theta+a^2\cos^2 
\theta)\right]\right\}\nn\\ 
B^{\hat\theta}=-\frac{1}{\Sigma\rho^4\sqrt{\Delta}}&&\left\{ 
B_{\|}\Delta\sin\theta 
\left[\rho^4r+a^2GM(r^2-a^2\cos^2\theta)(1+\cos^2\theta)\right]+\right.\nn\\&& 
\left. 
B_\bot\cos\theta\left[ 
\rho^4\left((\Delta r-GMa^2)\cos\psi+a(\Delta+GMr)\sin\psi\r)-\right.\right.\nn\\ 
&&\left.\left. 
a^2GMr\sin^2\theta\left(r^2(r-2GM)+2a^2(r\sin^2\theta+GM\cos^2\theta)\right)\cos\psi- 
\right.\right.\nn\\&&\left. \left.
a(\Delta-2GMr-2a^2\cos^2\theta)\sin\psi 
\right]\right\} 
\eeqa 
where  
\beq 
\psi=\phi+\frac{a}{2\sqrt{(GM)^2-a^2}}\ln\left[\frac{r-GM+\sqrt{(GM)^2-a^2}} 
{r-GM-\sqrt{(GM)^2-a^2}}\right]; \ \ B_\|=B_0\cos\chi,\ \ B_{\bot}=B_0\sin\chi. 
\eeq 
Rotation of the BH is responsible for the appearance of nonzero 
electric field whose  components are 
\beqa 
\label{E} 
E^{\hat r}=\frac{aGM}{\Sigma\Delta\rho^6}&&\left\{ 
B_{\|}\Delta\left[ 
2r^2\rho^4\sin^2\theta-(\Sigma^2-2GMra^2\sin^2\theta)(r^2-a^2\cos^2\theta) 
(1+\cos^2\theta)\right]-\nn\right.\nn\\&& \left. 
B_\bot r\sin\theta\cos\theta\left[ 
2\l((r\Delta-GMa^2)\cos\psi-a(\Delta+rGM)\sin\psi\r)+\right.\right.\nn\\&&\left.\left. 
(\Sigma^2-2GMra^2\sin^2\theta)\l(r^2(r-2GM)+2a^2(r\sin^2\theta+GM\cos^2\theta)\cos\psi- 
\right.\right.\right.\nn\\&&\left. \left.\left.
a(\Delta-2GMr-2a^2\cos^2\theta)\sin\psi\r) 
\right]\right\}\nn\\ 
E^{\hat\theta}=\frac{aGM}{\Sigma\sqrt{\Delta}\rho^6}&&\left\{ 
2B_{\|}r\sin\theta\cos\theta 
\left[\Delta\rho^4-(r^2-a^2)\l(\Sigma^2-2GMr(r^2+a^2)\r)\right]+\right.\nn\\ 
&&\left. 
B_\bot 
\left[ 
2r\rho^4\l(r(r\sin^2\theta+GM\cos 2\theta)\cos\psi- 
a(r\sin^2\theta+GM\cos^2\theta) 
\sin\psi\r)-\right.\right.\nn\\&&\left.\left. 
(r^2\cos 2\theta+a^2\cos^2\theta)\l(\Sigma^2-2GMr(r^2+a^2)\r)(a\sin\psi-r\cos\psi) 
\right]\right\} 
\eeqa 
 
\subsection{Equations of motion for a charged particle.} 
 
The components of the four-velocity of a particle $v^\mu=dx^\mu/dt$ in the  
orthonormal 
frame $e_{\hat a}$ (\ref{zamo1}) are  
\beq 
v^\mu=v^{\hat a}e^\mu_{\hat a}d\hat t/dt 
\eeq 
where $\hat t$ is the time which would be locally 
by the ZAMO observers  
at a given point and $t$ is the coordinate time which enters  
the metric (\ref{kerr}). For example, the $\phi$ components of particle  
velocity in coordinate and orthonormal reference frame are related through 
\beq 
v^{\hat\phi}=\frac{\Sigma\sin\theta}{\rho}\l(\frac{d\phi}{dt}- 
\frac{2GMar}{\Sigma^2}\r)\frac{dt}{d\hat t} 
\eeq 
The extra term $\Omega=2GMar/\Sigma^2$ is the angular velocity of the  
ZAMO  frame 
at each point. From (\ref{zamo1}) one can see that  
\beq 
\frac{d\hat t}{dt}=\frac{\rho\sqrt{\Delta}}{\Sigma} 
\eeq 
It is convenient to introduce a particle $\gamma$-factor in the orthonormal  
frame $\gamma=1/\sqrt{1- (v^{\hat a})^2}$.  
 
Equations of motion in the orthonormal basis (\ref{zamo})  have  the  
same form as in the flat space \citep{membrane} 
\beq 
\label{ham2} 
\frac{d{\vec p}}{d\hat t}= 
e(\vec E+\vec v\times \vec B)+m\gamma\vec g+\hat H\vec p+\vec f_{rad} 
\eeq 
where $\vec p$ is the particle momentum 
\beq 
\label{momentum} 
p^{\hat a}=m\gamma v^{\hat a} 
\eeq 
${\vec g}$ is the gravitational acceleration and ${\hat H}$ is the  
tensor of gravi-magnetic force. The force ${\vec f}_{rad}$ is the radiation 
reaction force. In the case of interest the time scales for acceleration in 
electromagnetic field and of the radiation reaction are orders of magnitude 
shorter than that of the motion of the particle in the gravitational field of the 
black hole. 
 
The radiation reaction force $\vec f_{rad}$  
for the ultra-relativistic particles moving in external electromagnetic field 
is (see, e.g. \citep{landau}) 
\beq 
\label{reaction} 
\vec f_{rad} 
=\frac{2e^4\gamma^2}{3m^2}\l((\vec E+\vec v\times \vec B)^2-(\vec v\cdot(\vec E+\vec v\times \vec B))^2\r)\frac{\vec v}{|v|} 
\eeq  
Note that if particles move at large angle with respect to the magnetic field 
lines, this expression will 
describe mostly synchrotron energy loss. However, in the case when particles 
move almost along the magnetic field lines, the last equation will "mimic" the 
effect of curvature energy loss (taking into account the fact that  
the typical curvature radius of the 
magnetic field lines in the considered case is about $R_{\rm Schw}$).


\begin{thebibliography}{} 
%
\bibitem[Aharonian(2000)] {synch_cut} Aharonian, F.A., 2000, New AR, 5, 377. 
%
\bibitem[Aharonian \& Neronov(2005)]{AhNer2005}  Aharonian F.A. and Neronov, A. 2006,  
Ap.J., 619, 306 
%
\bibitem[Aharonian et al.(2002)]{Aharonianetal2002} 	 
Aharonian, F. A., Belyanin, A. A., Derishev, E. V., Kocharovsky, V. V.,  
Kocharovsky, Vl. V. 2002, Phys ReV. D, 66, id. 023005 
%
\bibitem[Aharonian et al.(2003)] {HEGRA} 
Aharonian, F. et al. (HEGRA collaboration) 2003, A\&A, 403, L1 
%
\bibitem[Aharonian et al.(2006)] {HESS} Aharonian, F.  et al. (HESS collaboration) 2006, Science, 314, 1424 
%
\bibitem[Bardeen et al.(1972)]{bardeen} Bardeen, J.M., Press, W.H., 
Teukolsky, S.A., 1972, Ap.J. 178, 347. 
%
\bibitem[Beskin et al.(1992)]{beskin92} Beskin V.S., Istomin Ya.N., Par'ev V.I., 
 1992, Soviet Astronomy, 36, 642. 
%
\bibitem[Bicak et al.(1976)] {bicak} Bicak J., Dvorak L., 1976, Gen.Rel.Grav., 7, 959; see also 
Bicak J., Janis V., 1985, MNRAS, 212, 899. 
 
\bibitem[Biretta et al.(1999)]{biretta99} Biretta, J. A., Sparks, W. B., \& 
Macchetto, F., 1999, ApJ, 520, 621. 
 
\bibitem[Blandford \& Znajek(1977)]{blandford77} Blandford R.D., Znajek R.L., 
1977, MNRAS, 179, 433. 
 
%
\bibitem[Cheng et al.(1986)]{cheng86} Cheng K.S., Ho C., Ruderman M., 1986,  
 Ap.J. 300, 500. 
%
\bibitem[Cheung et al.(2007)]{cheung07} Cheung C.C., Harris D.E., Stawarz L., 
2007,  
 Ap.J.Lett. accepted, arXiv:0705.2448. 
 
\bibitem[Di Matteo et al.(2003)] {dimatteo03} 
Di Matteo T., Allen S.W., Fabian A.C., Wilson A.S. Young A.J. 2003, ApJ, 582, 
 133. 
 
\bibitem[Fabian(2006)] {Fabian} Fabian, A. 2006, Science, 314, 1398 
 
\bibitem[Georganopoulos et al.(2005)] {Markos}  
Georganopoulos, M., Perlman, E. S., Kazanas, D. 2005,  
ApJ, 634, L33 
%
\bibitem[Goldreich \& Julian(1969)] {goldreich} Goldreich P., Julian W.H.,  
1969, Ap.J. 157, 869. 
 
\bibitem[Harris et al.(2003)] {harris} Harris D.E., Biretta 
A.J., Junor W., Perlman E.S., Sparks W.B., Wilson A.S., 2003,  Ap.J.,  586, L41.  
 
\bibitem[Heinz \& Begelman(1997)]{heinz} Heinz S.; Begelman M.C., 1997, Ap.J., 490, 653 
%
\bibitem[Landau \& Lifshitz(1975)] {landau} Landau L.D., Lifshitz, E.M., 1975  
{\it The Classical Theory of Fields},  Oxford: Pergamon Press 
 
 
\bibitem[Levinson(2000)] {levinson} Levinson A., 2000, Phys. Rev. Lett., 85, 912. 
 
\bibitem[Marconi et al.(1997)] {marconi}  Marconi, A., Axon, D. J., Macchetto, F. D., Cappetti, A., Sparks, W. B.,  
Crane, P. 1997, \mnras, 289, L21  
%
 
\bibitem[Michel(2004)] {michel} Michel F.C., 2004, Ad.Sp.R., 33, 542. 
%
\bibitem[Neronov et al.(2002)] {neronov}  
Neronov A., Semikoz D., Aharonian F., Kalashev O., 2002, Phys.Rev. 
Lett., 89, 1101. 
 
\bibitem[Neronov et al.(2005)] {neronovtinyakov} Neronov A., Tinyakov P., Tkachev 
I., 2005, JETP, 100, 656. 
 
 
\bibitem[Perlman et al.(2001)] {perlman} Perlman E.S., Sparks W.B., Radomski J., Packham C., 
Fisher R.S., Pina R., Biretta J.A. 2001, Ap.J., 561, L51. 
%
%
\bibitem[Reimer et al.(2005)] {Anita} 
Reimer, A., Protheroe, R. J., Donea, A.-C. 2004, A\&A, 419, 89 
%
\bibitem[Sreekumar et al.(1994)] {sreekumar94} 
Sreekumar P. et al., 1994, Ap.J. 426, 105. 
%
\bibitem[Stawarz et al.(2005)] {Lukasz}  	 
Stawarz, L.,  Siemiginowska, A., Ostrowski, M.,  Sikora, M. 2005 
Ap.J., 626, 120  
%
\bibitem[Stawarz et al.(2006)] {HST-1}  
Stawarz, L., Aharonian, F., Kataoka, J., Ostrowski, M.,  
Siemiginowska, A., Sikora, M. 2006, MNRAS, 370, 981 
%
\bibitem[Thorne et al.(1986)] {membrane} Thorne K.S., Price R.H, Macdonald D.A.,  
1986, {\it Black 
Holes: the Membrane Paradigm}, Yale University Press. 
 
\bibitem[Tonry(1991)] {tonry91} Tonry, J. L. 1991, ApJ, 373, L1 
\bibitem[Wald(1974)] {wald} Wald R.M., 1974, Phys.Rev., D10, 1680. 
%
\bibitem[Whysong \& Antonucci(2004)] {whysong04} Whysong D., Antonucci R., 2004, 
Ap.J., 602, 16. 
%
\bibitem[Young et al.(2002)] {young02} Young A.J., Wilson A.S., Mundell C.G., 2002, Ap.J., 579, 560. 
 
\end{thebibliography}
\end{document}